\documentclass[nohyperref]{article}

\usepackage{todonotes, soul}

\usepackage{arxiv}

\usepackage[utf8]{inputenc} 
\usepackage[T1]{fontenc}    
\usepackage{hyperref}       
\usepackage{url}            
\usepackage{booktabs}       
\usepackage{amsfonts}       
\usepackage{nicefrac}       
\usepackage{microtype}      
\usepackage{lipsum}		
\usepackage{graphicx}
\usepackage{natbib}
\usepackage{doi}

\usepackage{graphicx}
\usepackage{subcaption}
\usepackage{placeins}

\usepackage{blkarray}      
\usepackage{multirow}
\usepackage{caption} 
\usepackage{array}
\newcommand{\PreserveBackslash}[1]{\let\temp=\\#1\let\\=\temp}
\newcolumntype{C}[1]{>{\PreserveBackslash\centering}p{#1}}

\usepackage{amsmath}
\usepackage{amssymb}
\usepackage{mathtools}
\usepackage{amsthm}
\usepackage{thmtools, thm-restate}

\usepackage{algorithmic}
\usepackage{algorithm}

\graphicspath{ {Figures/} }

\theoremstyle{plain}

\theoremstyle{definition}

\title{System architecture of a four-wheel drive Formula Student vehicle}

\date{} 					

\author{ \href{https://orcid.org/0000-0003-3412-0919}{\includegraphics[scale=0.06]{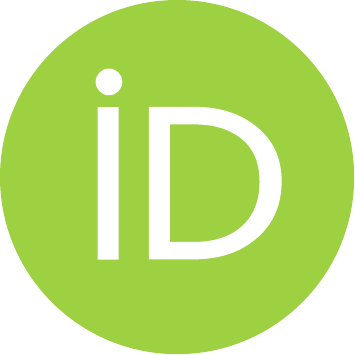}\hspace{1mm}Adriano~Schommer}\thanks{Corresponding author: adriano.schommer@brookes.ac.uk},
	\hspace{1mm}Gordana~Collier, 
	\hspace{1mm}Robert~Norris, 
	\hspace{1mm}Denise~Morrey,
	\hspace{1mm}Ludmila N.~Maria \\
	Faculty of Technology, Design and Environment\\
	Oxford Brookes University\\
	OX33 1HX, Oxford, UK \\
	\And
	Chris~Johnston \\
	Jaguar Land Rover \\
	Vehicle Dynamics \& Digital Engineering Group \\
	CV35 0RR, Warwick, UK \\
}

\renewcommand{\shorttitle}{System architecture of a four-wheel drive Formula Student vehicle}

\begin{document}
\maketitle

\begin{abstract}
Formula Student vehicles are becoming increasingly complex, especially
with the current shift from internal combustion engines toward electric
powertrains. The interaction between software and hardware is complex
and imposes additional challenges for systems integration. This paper
provides a structured introduction to the OBR22 Oxford Brookes Racing
Formula Student electric vehicle. From a system architecture perspective,
the four-wheel drive in-hub motors topology is described. Diagrams
of the hardware components, the architecture of the high voltage and
communication systems are presented. This paper also demonstrates
the model-based development process, including an overview of the
model-in-the-loop (MiL) and hardware-in-the-loop (HiL) control design
phases.
\end{abstract}


\section{Introduction}

The fast-paced and high turnover rate of a typical Formula Student
team highlights the importance of knowledge transfer for long-term
success. In Oxford Brookes Racing, most team members contribute for
a single year during their Master of Science courses, which results
in a season-to-season turnover rate as high as 80\%. To attenuate
this challenge, the development process of vehicle systems should
be straightforward. And this is where automotive industry practices
can be valuable, provided the core principles are adapted to the needs
and limitations of the Formula Student environment.

With the ever-increasing complexity of Formula Student vehicles, especially
concerning the electric tractive system, the boundaries between systems
have become blurred. Hardware, software, vehicle controls design and
vehicle dynamics design overlap in a way that they no longer can be
considered individual parts. Instead, they are seen as different perspectives
of the same system \citep{Galbraith2019}. A consistent vehicle development
process is needed for the Formula Student project, capable of improving
both design agility and continuity from season to season. 

To support new team members, two areas of improvement were identified:
firstly, a better understanding of the system architecture, and secondly,
clarity on how engineering sections and tools should interact. To
address the first issue, a formal documented overview of the system
architecture is a good starting point, this approach was demonstrated
in a complex autonomous system for a driverless Formula Student vehicle
\citep{Kabzan2020}. To deal with the second issue, a road map of
the design process must be provided. When team members have a clear
picture of how the vehicle is developed, the interaction between engineering
sections becomes easier to understand.

 To address project complexity, a methodology should consider requirements
and interactions between subsystems early in the system design. A
relevant approach is model-based systems engineering (MBSE). It takes
the systems engineering premises - defining what a system must do,
how well a system must work, and how system functions are tested \citep{Richards2020},
using modelling and simulation as the means of information exchange
instead of a document-centric approach \citep{Capasso2017}. The primary
focus of MBSE is to ensure integration and understanding of a system
from multiple perspectives \citep{Grssler2021}. This process is guided
by a V-model framework that is represented by the design process from
requirements on the left to hardware integration. The V-model approach
is further explored in section \ref{sec:5}.

\begin{figure}[h]
\begin{centering}
\includegraphics[width=12cm]{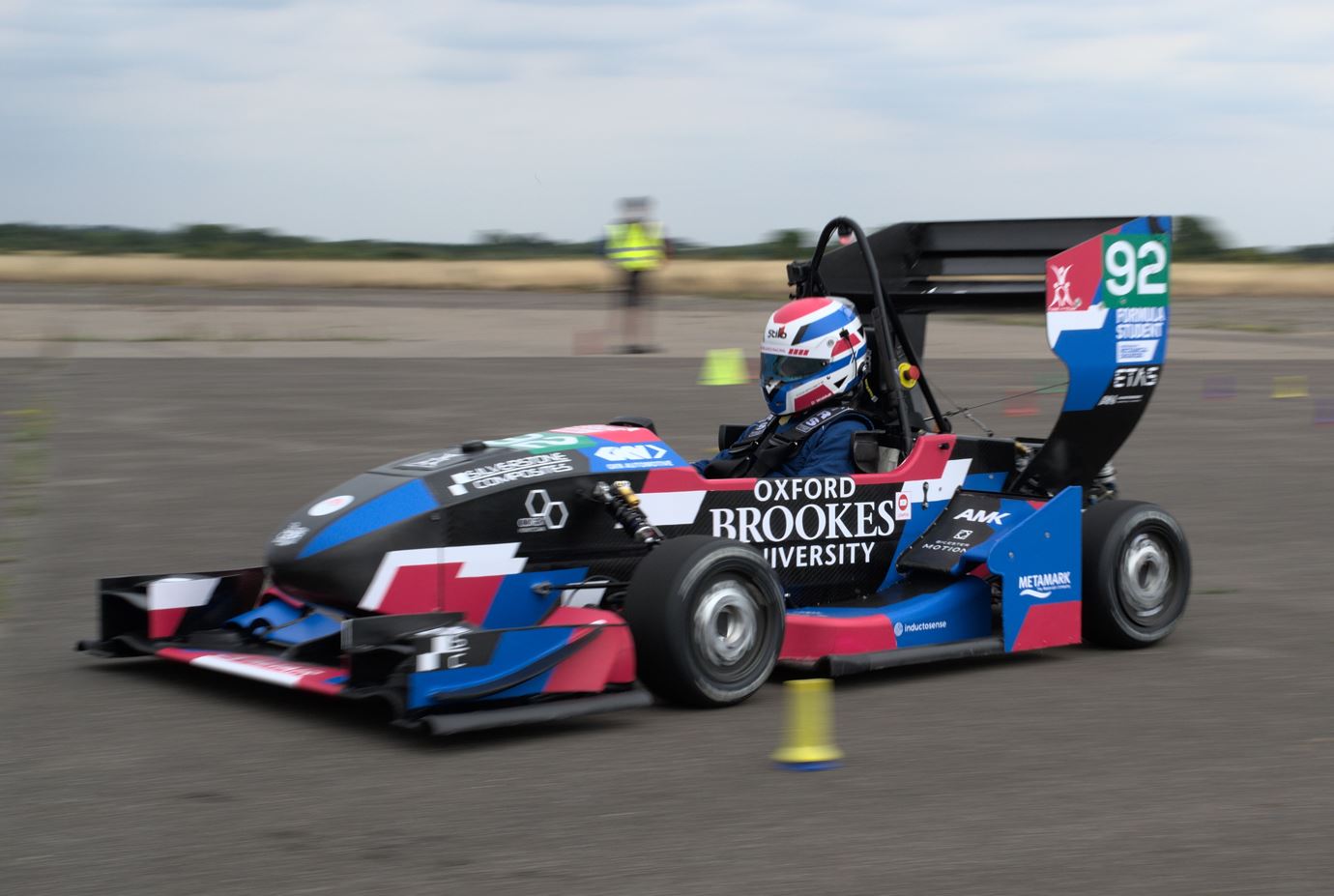}
\par\end{centering}
\caption{OBR22 Formula Student electric vehicle.}
\label{figure_OBR22}
\end{figure}

In the context of a Formula Student team, a V-model development process
was demonstrated by \citep{Schumacher2021} in the design of a four-wheel
drive (4WD) electric drivetrain, from concept to system realization.
In this MBSE approach, an executable platform was created using AVL
eSUITE™ software to run multi-domain simulations, which enabled the
design and integration of components along the entire design cycle.
From a mechanical design viewpoint, the concept of the in-hub motor
was also addressed by \citep{Kucinski2017} from top-level requirements
to hardware validation. However, the interfaces between software and
hardware in the development process of a Formula Student electric
vehicle are still not clear from the literature. This paper address
this gap by providing an overview of the system architecture and development
process of the OBR22 electric powertrain, the first electric vehicle
from the Oxford Brookes Racing team (Figure \ref{figure_OBR22}). 

In the following sections, the OBR22 4WD system is presented with
increasing levels of detail from different points of view. Firstly,
software and hardware development aspects are distinguished in Section
\ref{sec:2}, followed by a brief description of the battery pack,
motors and inverters. Sections \ref{sec:3} and \ref{sec:4} present
the high voltage and communication systems architecture. Finally,
section \ref{sec:5} gives an overview of the MBSE methodology, and
explores the development process from a model-in-the-loop (MIL) and
hardware-in-the-loop (HIL) perspectives. Section \ref{sec:6} concludes
this paper.

\section{System Overview\label{sec:2}}

\subsection{Vehicle Control Unit (VCU): the interface between software and hardware}

The majority of the Formula Student vehicles running internal combustion
engines use off the shelf programmable electronic control units (ECU)
for engine management. Although it is possible to develop new advanced
algorithms, usually programmable engine ECUs are only required to
be configured and calibrated to a particular engine design. For electric
vehicles however, despite of having fewer components compared to an
engine vehicle, the number of different vehicle system typologies
requires customized software control solutions to be closely related
to the system hardware \citep{Schumacher2021,Badal2019}. 

In a 4WD in-hub architecture, the management of the torque request
for each motor is crucial for achieving a functioning system. This
alone requires development of specialized vehicle control algorithms
as an inherent part of the tractive system. It is needed for performance
features and also for the safety limits imposed by competition rules
\citep{fsrules2022}. Therefore, the development of control algorithms
for a 4WD in-hub Formula Student vehicle is not a choice, it is very
much part of the system, in the same way as the battery cells are
or any other hardware component.

\begin{figure}[h]
\begin{centering}
\includegraphics[width=7.5cm]{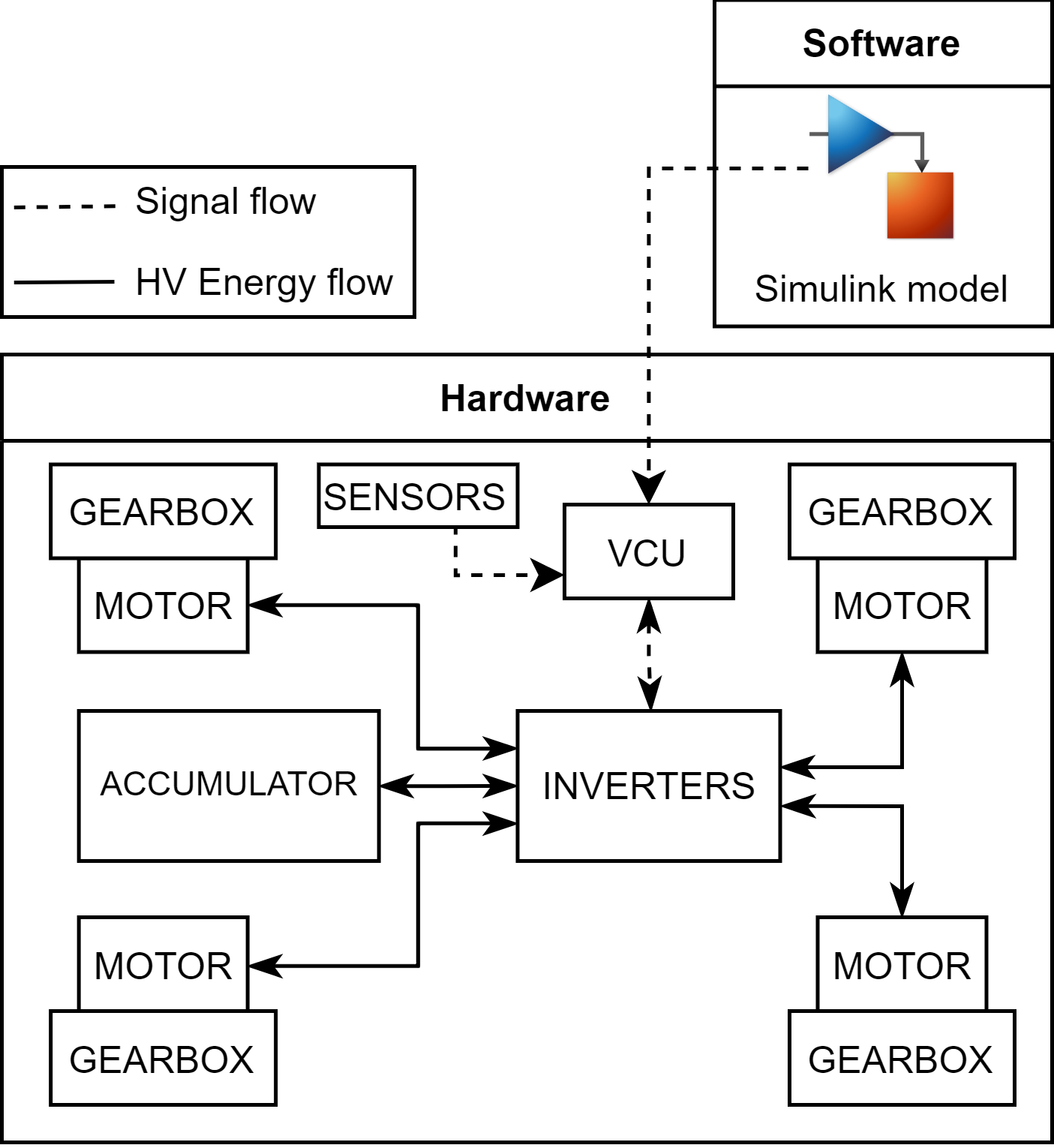}
\par\end{centering}
\caption{Hardware and software overview of the electric vehicle. Software is
developed in Simulink and uploaded into the VCU, which is the interface
between software and hardware. The VCU takes inputs from sensors and
sends a torque request to the inverters. The accumulator is the high
voltage supply of the system.}
\label{figure_hardware_software}
\end{figure}

Figure \ref{figure_hardware_software} illustrates this concept, where
software and hardware are highlighted as two major areas of the OBR22
powertrain development. From a simplified standpoint, the algorithm
developed in Simulink \citep{Simulink2022} controls the energy flows
within the tractive system: it provides limits for power output and
manages the torque requested by the driver across all four wheels
based on vehicle sensor readings.

The vehicle control unit (VCU) is a supervisory controller that provides
the top level control between all the vehicle subsystems. Where each
vehicle subsystem is typically controlled by a local electronic control
unit (ECU). A strict differentiation between VCU and ECU devices varies
among manufactures, however in general terms a VCU due to its flexible
nature is typically a fully programmable real-time computer.

The VCU in the OBR22 vehicle is an ETAS ES910 rapid prototyping module,
a real-time computer with significantly higher computing performance
than an ECU \citep{etas910}. Typically VCUs provide an integration
with a platform such as MATLAB/Simulink to enable the development
of signal flows and state machines, and this is how the OBR vehicle
controls algorithms are developed, with the help of INTECRIO software
\citep{intecrio2022} and its INTECRIO-RLINK Simulink Blockset \citep{intecrioRLINK2022}.

\subsection{Motors and inverters}

From a hardware standpoint, the main components of an electric powertrain
are: motors, inverters, accumulator and gearbox. OBR22 uses off-the-shelf
motors and inverters provided by AMK in the package "Racing Kit
4 wheel drive Formula Student Electric". A summary of
the motor AMK DD5-14-10-POW technical data is given in Table 1, where
the rated operating condition relates to continuous output with controlled
heat dissipation and the maximum operation relates to output over
a short time that avoids overheating the motor windings. The AMK inverters
limit the maximum operation condition to a duration of 1.24 seconds.

\begin{table}[h]
\centering{}\caption{Summary of the motor AMK DD5-14-10-POW technical data \citep{amk2020}.}
\begin{tabular}{>{\centering}b{3cm}>{\centering}p{3cm}>{\centering}p{3cm}}
\toprule 
\textbf{Parameter} & \textbf{Rated Value} & \textbf{Maximum Value}\tabularnewline
\midrule 
Power & $12.3$ $kW$ & $35^{1}$$kW$\tabularnewline
\midrule 
Torque & $9.8$ $N.m$ & $2$1 $N.m$\tabularnewline
\midrule 
Current & $41^{*}$ $A$ & $105^{*}$ $A$\tabularnewline
\midrule 
Speed & $12000$ $rpm$ & $20000$ $rpm$\tabularnewline
\midrule 
\multicolumn{3}{l}{$^{1}$ for a voltage supply of 600 VDC}\tabularnewline
\multicolumn{3}{l}{$^{*}$ root mean squared current draw}\tabularnewline
\end{tabular}\label{table_amk_motor_technical_data}
\end{table}

The AMK motor is a permanent-magnet synchronous motor (PMSM). Compared
to an induction motor, a PMSM design with an equivalent power rating
have increased power density, greater torque-to-inertia ratio, are
more efficient and easier to cool even though PMSM motors are more
sensitive to higher operating temperatures \citep{husain2021electric}.
Although the peak power of each motor is 35kW (as shown in Table \ref{table_amk_motor_technical_data}),
the total maximum power of the powertrain is limited to 80kW by the
competition rules \citep{fsrules2022}. In addition, the maximum power
is dependent of the supply voltage from the accumulator, which decreases
as the state of charge of depletes. This effect was further explored
in \citep{Barham2017}, where the effect of various supply voltages
on power and torque were analysed.

The AMK motor operates with alternate current (AC), whereas the battery
pack supplies direct current (DC). The primary function of the inverters
is to convert the battery DC into motor AC. This is accomplished by
using high frequency switches to activate the current flow through
the windings of the motors in a synchronous manner. Each inverter
uses six insulated-gate bipolar transistors (IGBTs). Compared to alternatives
such as silicon carbide (SiC) MOSFETs, these high frequency switches
are considered to be the main disadvantage of the AMK package due
to its size, weight and its lower efficiency \citep{Galbraith2019}.
For more information on how the inverters work the reader is referred
to \citep{larminie2012electric}.

\begin{figure}[h]
\begin{centering}
\includegraphics[width=9.5cm]{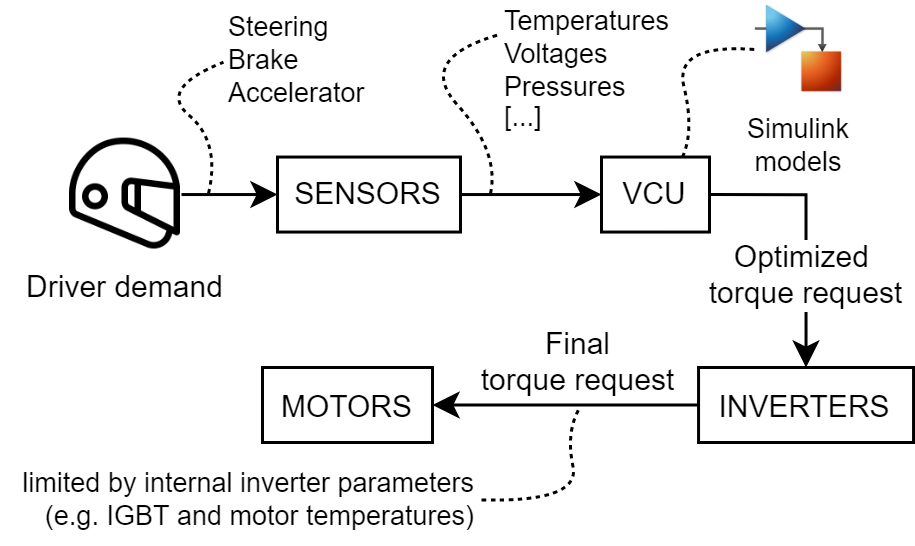}
\par\end{centering}
\caption{Energy flow of torque request. Inputs from the driver are translated
by sensors and sent to the VCU where an optimized torque request is
calculated based on vehicle sensors. This torque request is sent to
the inverters where it is checked against internal limitations and
DC current is converted into AC current to power the motors.}
\label{figure_energy_flow}
\end{figure}

Figure \ref{figure_energy_flow} summarizes the energy control flow
between the VCU, motors and inverters.This starts with driver's input
(accelerator pedal position, brake pressure and steering angle) and
the vehicle sensor readings, from which the VCU algorithms evaluate
the vehicle states and imposes limits for maximum performance (e.g.
torque vectoring, energy consumption). Then, the VCU sends to the
inverters an optimized torque request for each wheel, where another
check of safety is performed by the inverters to make sure the temperatures
of the motors and IGBTs are within safe limits. Finally, the current
is controlled by the inverters and switched through its IGBTs to provide
AC current to the motors where electrical energy is converted into
mechanical energy.

\subsection{Accumulator}

\begin{figure}[h]
\begin{centering}
\includegraphics[width=11cm]{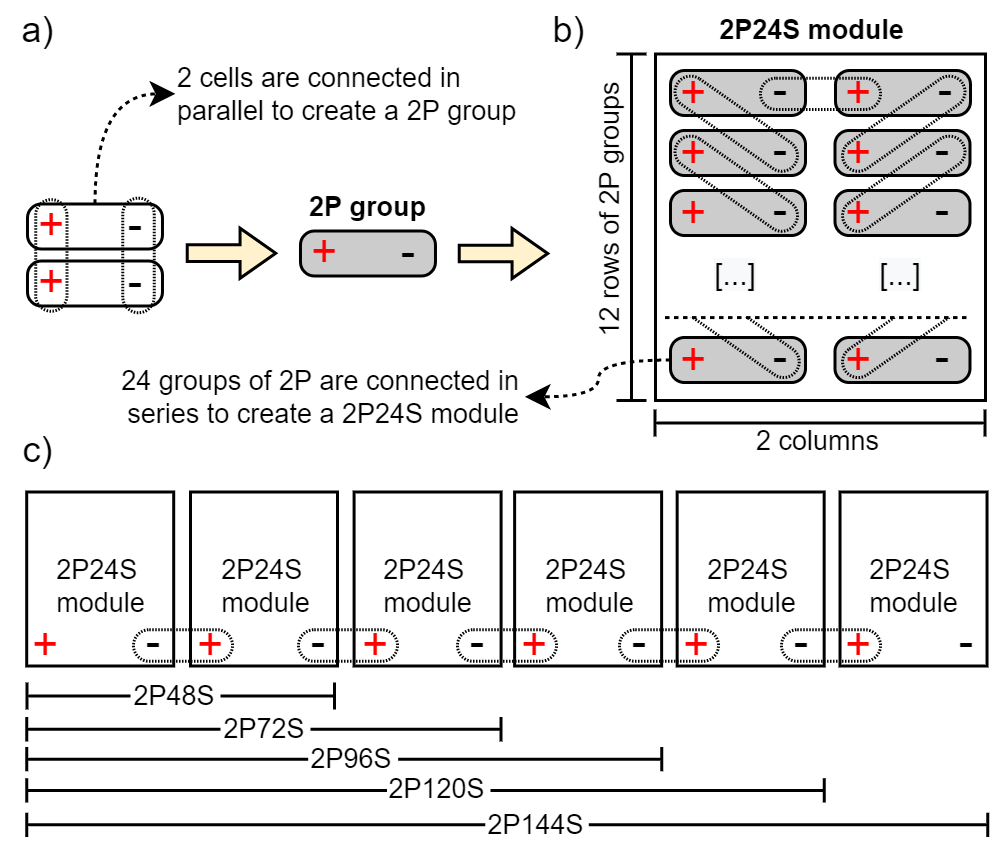}
\par\end{centering}
\caption{OBR22 Accumulator. a) cells are grouped in pairs to create a 2P (two
parallel) connection; b) 24 groups of 2P cells are further connected
series to create a 2P24S module; c) six modules are then connected
in series to form the 2P144S accumulator.}
\label{figure_accumulator}
\end{figure}

There is a great variety of lithium-ion battery cells in the market.
From cell chemistry to the form factor, each cell has its unique characteristics
and trade offs. There are three common types of form factors: cylindrical,
prismatic and pouch cells. Both cylindrical and prismatic are contained
in hard cases, whereas pouch cells are encased in soft pouches. Due
to the dynamic pressure during charge and discharge of lithium-ion
battery cells, pouch cells require a constant external pressure to
increase mechanical stability. This pressure is directly related to
performance aspects (for example, aging and capacity fade), and safety
aspects such as dynamic instability due to gas formation, a problem
known as swelling \citep{Zhou2020}. Despite these design requirements,
pouch cells have higher power density $[W/m^{3}]$ and energy density
$[Wh/m^{3}]$ than cylindrical and prismatic cells. Additional, they
can be packed more closely, saving space and reducing mass. 

\begin{table}[h]
\caption{Overview of the Melasta SLPBA442124 cell \citep{melasta2019}.}

\begin{centering}
\begin{tabular}{>{\centering}p{5cm}>{\centering}p{3cm}>{\centering}p{3cm}}
\toprule 
\textbf{Parameter} & \textbf{Discharge} & \textbf{Charge}\tabularnewline
\midrule 
Maximum continuous current & 110 $A$ & 11 $A$\tabularnewline
\midrule 
Peak Current ($\le3s$) & 137.5 $A$ & 16.5 $A$\tabularnewline
\midrule 
Voltage & 3 $V$ & 4.2 $V$\tabularnewline
\midrule 
Operating temperature & -20 \textasciitilde 60 ${^\circ}C$ & 10 \textasciitilde 45 ${^\circ}C$\tabularnewline
\midrule 
\multicolumn{1}{c}{Nominal Capacity} & \multicolumn{2}{c}{5.5 $Ah$}\tabularnewline
\midrule 
Weight & \multicolumn{2}{c}{116 $g$}\tabularnewline
\end{tabular}
\par\end{centering}
\label{table_melasta_cell_technical_data}
\end{table}

The cell chemistry is beyond the scope of this paper, but the technical
data of OBR22 Lithium Cobalt Oxide (LCO) pouch cells is presented
in Table \ref{table_melasta_cell_technical_data}. The maximum discharge
capacity that a cell can safely deliver is described as the C-rate,
computed as the maximum continuous current {[}$A${]} divided by its
nominal capacity {[}$Ah${]}. The Melasta SLPBA442124 has a C-rate
of 20, which in practice means the cell could be discharged entirely
in 3 minutes. For further information about battery terminology the
reader is refereed to \citep{mev2008guide}.

Besides the battery cell, the battery pack configuration is an important
aspect of the accumulator design. A combination of series and parallel
arrangement is normally used to meet voltage and current requirements.
According to the competition rules, any connection between two electrical
components of the tractive system must not exceed the voltage of 600V
(EV4.1.1 \citep{fsrules2022}). Within this constraint, higher voltages
are preferred because power losses ($P_{loss}[W]$), computed as $R.I^{2}$,
are proportional to the square of the current $I[A]$ and proportional
to the electrical resistance of the system $R[\varOmega]$. Therefore,
for the same amount of power output $P[W]$, higher voltages require
lower current draw because $P=V.I$, and consequently less power loss,
smaller wire gauges and ultimately lighter components. The OBR22 accumulator
(Figure \ref{figure_accumulator}) operates with $600V$ at full state
of charge and contains 288 cells arranged in a 2P144S configuration
(two cells connected in parallel for each of the 144 connection in
series). 

To comply with the rule EV 5.3.2 \citep{fsrules2022}, which limits
the accumulator segments to a maximum of 6MJ and 120V DC, the 2P144S
configuration is further divided into six segments wired in series,
each in a 2P24S configuration (48 cells configured with two cells
connected in parallel and 24 cells in series). The accumulator has
a nominal energy capacity of $5.8kWh$, computed using Equation \ref{eq:equation_accumulator_capacity}
\citep{plett2015battery},

\begin{equation}
Q_{pack,nom}=N_{cells}.V_{nom}.Q_{cell,nom}\label{eq:equation_accumulator_capacity}
\end{equation}

where $N_{cells}$ is the total number of cells in the accumulator,
$V_{nom}[V]$ is the nominal cell voltage and $Q_{cell,nom}[Ah]$
is the nominal cell charge capacity.

\section{High Voltage system architecture\label{sec:3}}

\begin{figure}[h]
\begin{centering}
\includegraphics[width=16cm]{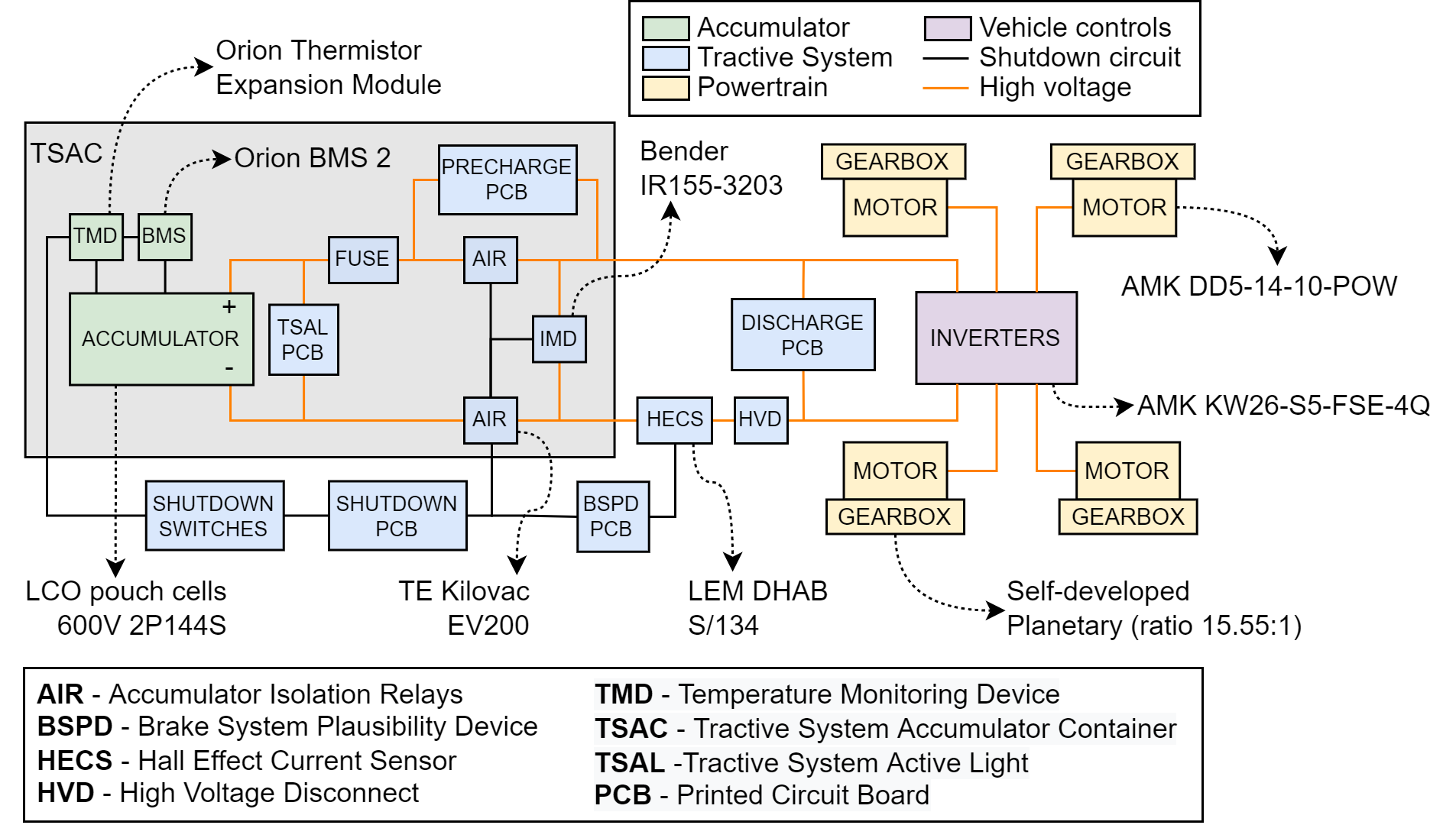}
\par\end{centering}
\caption{OBR22 high voltage system architecture.}
\label{figure_HV_overview}
\end{figure}

The OBR22 high voltage system comprises motors, accumulator, the tractive
system -- defined by the rules as every part electrically connected
to the motors and the accumulator, and several safety features connected
to a centralized shutdown circuit. Figure \ref{figure_HV_overview}
gives an overview of the high voltage system and how the development
areas are structured within OBR team.

The Tractive System Accumulator Container (TSAC) is surrounded by
strict rules and encloses all unswitched high voltage components.
Its development includes notable integration of many areas of development
such as composites -- for mechanical design and manufacturing of
the casing, tractive system, low voltage electronics and electrical
integration. The Battery Management System (BMS) within the TSAC is
responsible for estimating the battery states (for example the state
of charge and state of health). The monitoring of cell voltages and
temperatures is performed in OBR22 by an auxiliary Temperature Monitoring
Device (TMD). The BMS is connected to the shutdown circuit and can
open the two Accumulator Isolation Relays (AIR) in case any safety
limit is exceeded. Opening the AIRs completely shuts the high voltage
supply from the TSAC to the vehicle. 

The high voltage system also includes printed circuit boards (PCBs)
specifically designed to ensure overall system compliance: The Tractive
System Active Light (TSAL) PCB indicates the status of the system
and the Brake System Plausibility Device (BSPD) PCB checks if power
is being applied to the motors whilst the vehicle is under hard braking,
if true it opens the shutdown circuit through the shutdown PCB. This
is done by measuring the HV current via a Hall Effect Current Sensor
(HECS) and a brake pressure sensor. The precharge PCB protects the
inverter capacitors from inrush current from the accumulator when
the AIRs are closed, whereas the discharge PCB removes potential dangerous
residual high voltage stored in the capacitors after the AIRs are
opened. Finally, the last component depicted in the diagram, the High
Voltage Disconnect (HVD), is a removable connection of the accumulator
negative pole. By rules, the HVD must be manually removable by an
untrained person without the need of tools. 

\section{Communication architecture\label{sec:4}}

\begin{figure}[h]
\begin{centering}
\includegraphics[width=16cm]{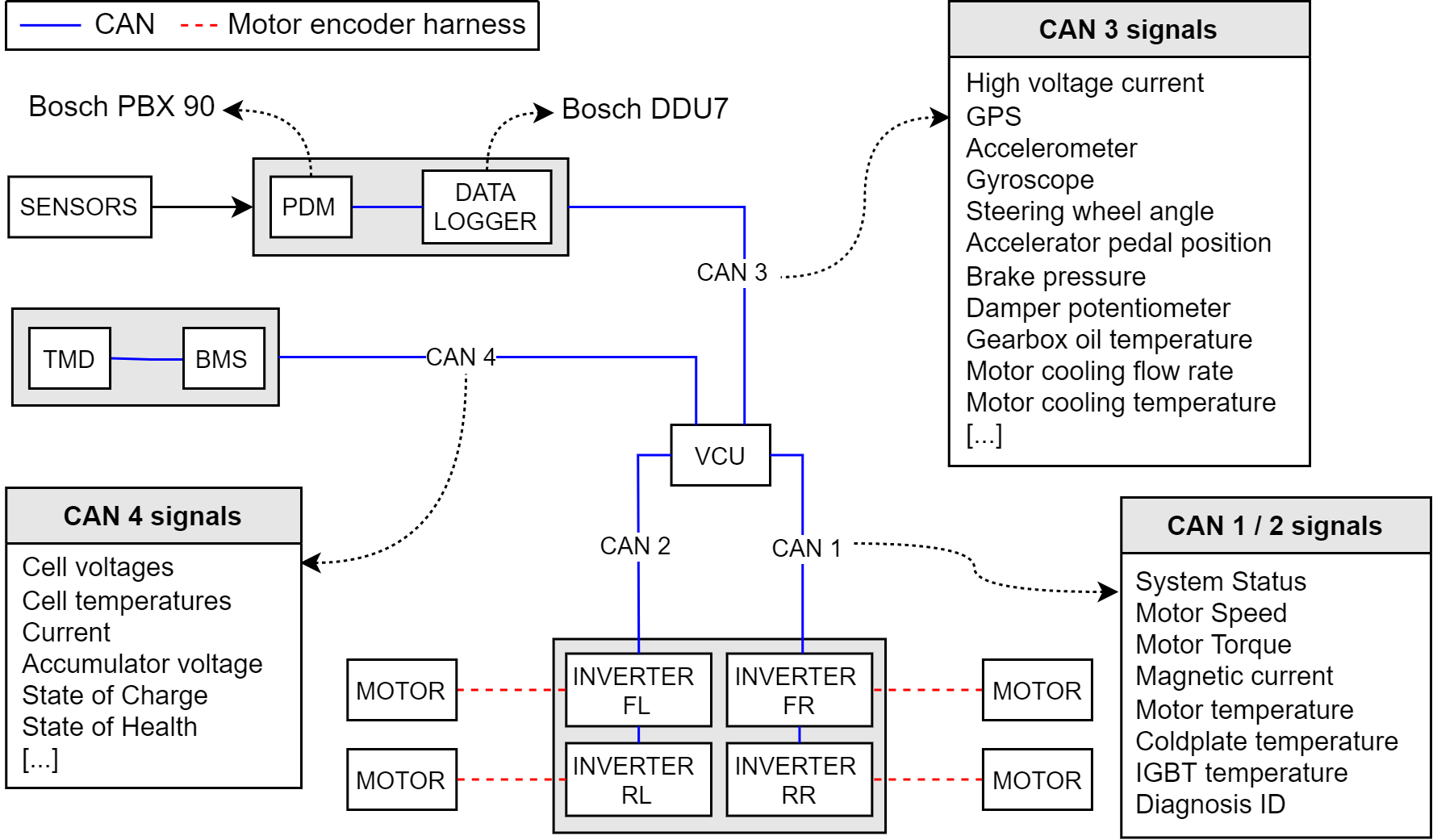}
\par\end{centering}
\caption{OBR22 vehicle communication system architecture.}
\label{figure_communications_view}
\end{figure}

The backbone of OBR communication is the Controller Area Network (CAN).
Channels from sensors and devices are integrated by four main CAN
bus lines at vehicle level (Figure \ref{figure_communications_view}). 

CAN 1 and 2 connect the inverter controllers with the vehicle control
algorithms to the VCU. Motor speed, torque, temperature are sent to
the inverters by encoders mounted to each motor, which are sent via
CAN by the inverters to the VCU and to the data logger where they
are recorded. The vehicle sensors are integrated in CAN 3, they are
either powered by the Power Distribution Module (PDM) or directly
connected to the data logger. The PDM is a power distribution unit
that replaces conventional relays and fuses, and the data logger is
integrated into the Bosch DDU7 dashboard display. Finally, the accumulator
has its own CAN 4 with several configurable channels, including voltages
and temperatures at cell and battery pack level. 

\subsection{Configuration software}

\begin{figure}[h]
\begin{centering}
\includegraphics[width=15cm]{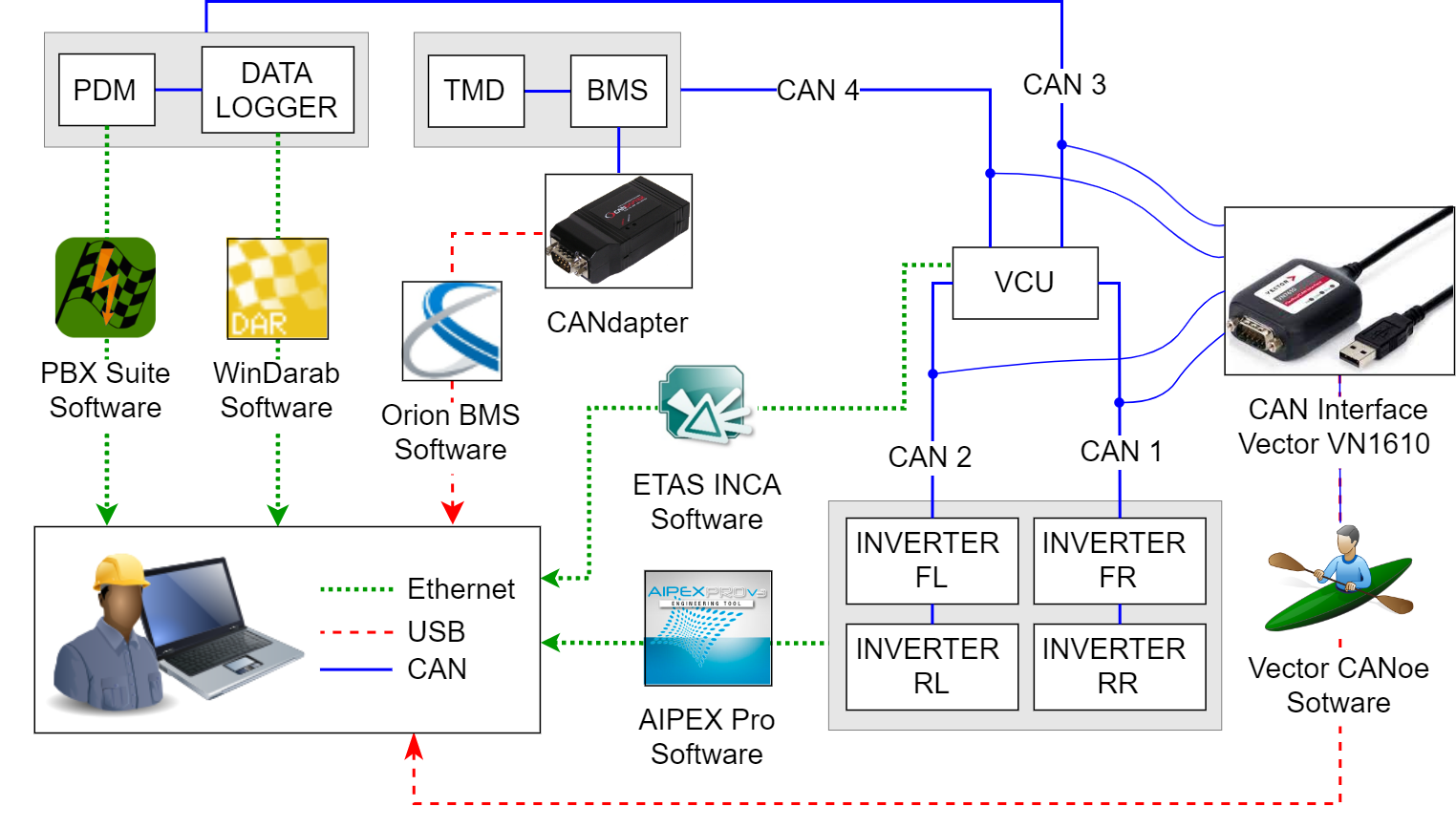}
\par\end{centering}
\caption{OBR22 user-vehicle communication system architecture.}
\label{figure_communications_view_user_vehicle}
\end{figure}

Figure \ref{figure_communications_view_user_vehicle} illustrates
the software used with each device and the type of connection with
PC (USB or Ethernet cable). This software allows the configuration
of devices, and access to data on CAN buses. CAN signals can also
be accessed directly via USB connection with the help of a CAN interface
\citep{vectorCANinterface} and the Vector CANoe software \citep{vectorCANoe}
(right side of the diagram). This tool is tailored to deal with CAN
buses and enables both reading and sending signals to devices.

The VCU is accessed throughout ETAS INCA Software \citep{etasINCA}
via Ethernet connection and it is the means of loading the vehicle
controls algorithms into the VCU. The inverters are accessed via Ethernet
cable using the AMK AIPEX Pro software \citep{amk2019aipexpro}, each
inverter has its own Ethernet port and therefore might be connected
one at a time in a PC. The PDM module is accessed through Ethernet
and is configured with the help of the Bosch PBX Suite software. Finally,
the BMS needs a CAN adapter to be accessed through the Orion BMS software
in which the accumulator topology and the CAN 4 channels are configured.

These controller devices are extensively accessed during design phases,
whereas during vehicle testing phase CAN channels are centralized
in the data logger software Bosch WinDarab \citep{bosch2021windarab}.
This is a data acquisition and analysis tool capable of displaying
signals in various formats including plots over time or distance,
helpful to identify trends and issues because the correlation to other
physical, temporal and location measures can be inferred all in a
single tool (e.g. temperatures, bumps in the tarmac, high acceleration
corners).

\section{Model-based approach for vehicle controls development\label{sec:5}
}

Because the competition event has a predefined timeline, project planning
quickly becomes a major factor of success. Testing time on track is
valuable and sensible project planning is needed but challenging to
achieve considering the high turnover rate of team members. Therefore,
the development process must allow new engineers to quickly understand
the working principles of the system, how design phases are structured
and essentially what needs to be accomplished in each phase of the
project. 

New designs naturally involve high levels of uncertainty, however
most of the time systems are not designed from scratch, but rather
are an evolution from a previous design. The project plan then becomes
a task list of what needs changing for each design iteration and what
is already known to work well is left alone. The ultimate goal is
to achieve continuous development where there is a fast iteration
from one stable design to the next stable design. 

\begin{figure}[h]
\begin{centering}
\includegraphics[width=17cm]{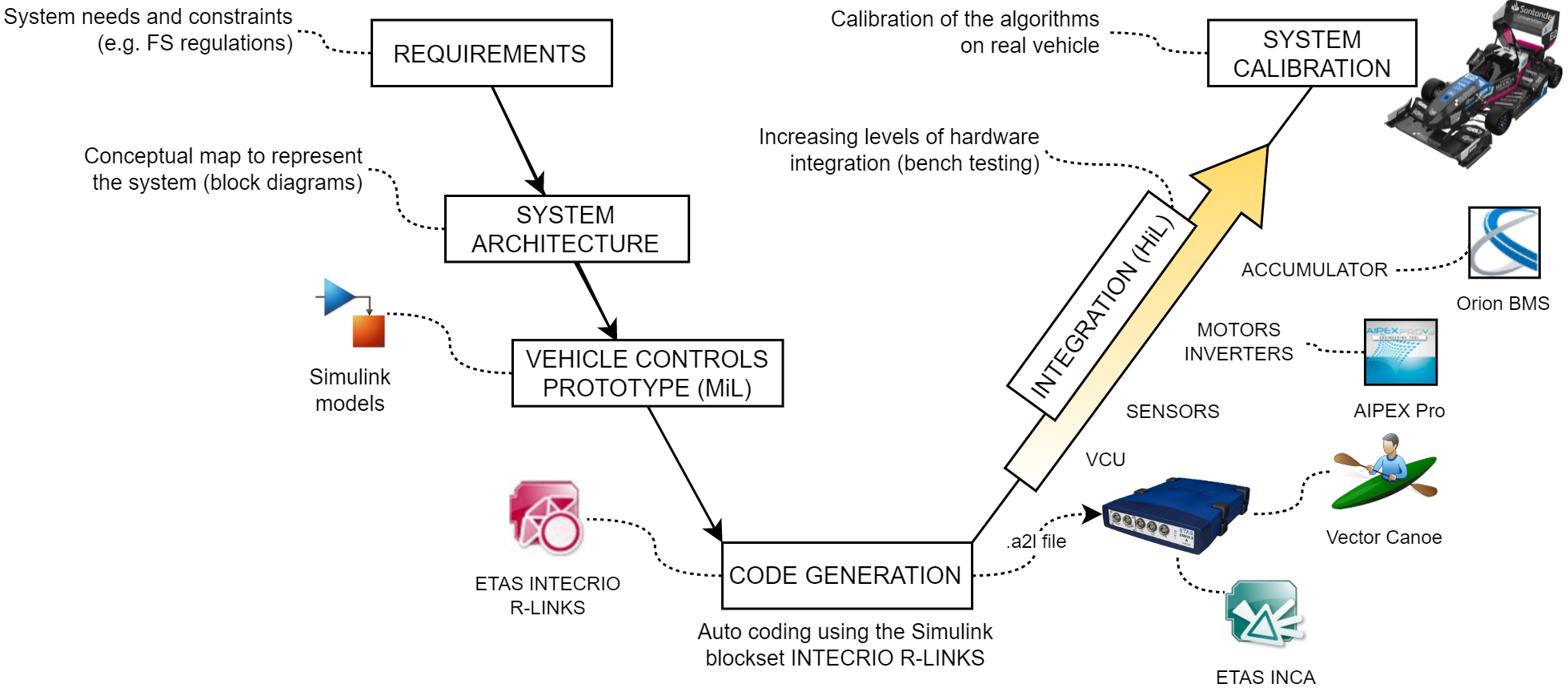}
\par\end{centering}
\caption{V-model development process.}
\label{figure_v-model}
\end{figure}

Figure \ref{figure_v-model} gives an overview of the OBR model-based
development process for vehicle controls and illustrates the various
software tools used in each phase of the project. On the left side
of the diagram, the objective is to propagate design requirements
from a system to a component level in a top-down approach. On the
right side, the objective is to validate the system performance by
increasing integration of components in a bottom-up approach \citep{Capasso2017}.
To further explore the V-model, an overview of each phase of the process
is given, starting from the requirements on the top-left of the diagram. 

\subsection{Requirements and system architecture}

The system design is constrained by technical requirements and competition
regulations. The documentation of these requirements is the first
phase of the design process, it helps understanding the reasoning
behind design decisions and the critical aspects new developments
that should be taken into account. In fact, it is not effective to
start developing new functionalities if the existing system is not
reasonably understood. There is a learning curve, that takes a short
time or long time depending on the complexity of the system, however,
experience shows that comprehending the top-level working principles
of the system and the relationship between components makes the lower
level technical details easier to follow as consequence. The second
stage of the process, system architecture, addresses this gap (that
is figures \ref{figure_HV_overview}, \ref{figure_communications_view}
and \ref{figure_communications_view_user_vehicle}) by representing
the system from different perspectives, to provide a conceptual map
of the system.

\subsection{Vehicle controls prototype (MiL) and code generation}

The development of vehicle control algorithms happens in the third
phase: model-in-the-loop. Vehicle controls can be thought as a black
box input/output system which takes sensors inputs (steering angle,
accelerator and brake pedal positions, for example) and transforms
it into a torque request to each motor. These algorithms are developed
in Simulink and tested in a virtual environment before they are implemented
in the real vehicle. This approach has several advantages, engineers
can conduct tests at critical conditions that are not possible or
difficult to be safely replicated in real life. For example, the torque
limitation to prevent battery cells of exceeding critical temperatures
can be tested multiple times to find the optimum strategy to increase
performance while keeping the system safe.

Once the vehicle control algorithms are modelled and tested in Simulink,
they need to be uploaded into the VCU. However, to make best use of
the computational power of the VCU, Simulink models are not deployed
directly into it. Instead, efficient code is generated with the help
of a proprietary Simulink Blockset ETAS INTECRIO R-LINK \citep{intecrioRLINK2022}.
This creates an executable file (.a2l) which is then uploaded via
an ethernet connection using the ETAS INCA software. Once this process
is completed (a couple of minutes), the VCU is ready to run. 

\subsection{Integration (HiL) and system calibration}

Ideally, the virtual vehicle represents the real vehicle very closely,
however, vehicle models are only a representation of the real system
and deviations from reality are an intrinsic part of the process.
During this phase the actual vehicle is still not finished and the
physical properties that cannot be measured need to be estimated.
The integration of real hardware attempts to fill the gaps between
the virtual and real environment using a variety of bench testing.

A detailed description of such tests are beyond the scope of this
paper and largely depends on the specifics of the project, which changes
from year to year. Nevertheless, in general terms the integration
begins with the VCU and, one step at time, sensors are included in
the system and monitored in a PC with the help of the Vector CANoe
software, capable of displaying information from any of the four CAN
buses. Next, the motors and inverters are integrated and monitored
using the AMK AIPEX Pro software. At this stage the inverters are
powered by a high voltage power supply which is later substituted
by the accumulator and monitored with the Orion BMS software. 

The integration phase aims first to check basic functionalities such
as CAN communication, and to test if the physical behaviours of the
components are as expected. At the same time, this creates clear milestones
that can be restored if problems occur in the following phases of
integration. This is the reason for progressively incorporating hardware
components instead of connecting the entire system all at once. It
is easier to find and prove out each localised subsystem at time.
This process continues until the entire system is at its full functioning
state. Finally, there is a transition from bench testing to the real
vehicle testing: the system calibration phase. This stage deals with
the final details of hardware implementation and aims to fine tune
the vehicle control algorithms. In reality, the V-model is not a linear
process tasks cycling back and forth with iterations between the integration
and calibration phases, and with the vehicle controls prototype. This
iterative approach also feeds back to the requirements that are needed
to achieve a better and safer design. 

\section{Concluding remarks\label{sec:6} }

 The system architecture of a Formula Student vehicle with 4WD in-hub
motors was introduced. First, the interface between hardware and software
was demonstrated. Major components of the electric system such as
the accumulator, inverters and motors were described and the interfaces
with other devices of the high voltage system were illustrated in
the format of block diagrams. In a similar manner, the communication
architecture was presented from a vehicle perspective and a user perspective.
Software tools supporting engineers to access the electronic devices
and data on the CAN buses clarified when and how each tool is used
along project phases. Finally, the OBR vehicle controls V-model development
process was presented. This methodology was based on the current automotive
industry approach but simplified to promote smooth integration of
young engineers.

Designing a Formula Student electric vehicle faces similar challenges
to the challenges that the automotive industry faces in the race towards
electric mobility. It stretches the limits of knowledge, and the efficient
usage of resources and promotes intense teamwork. The collaboration
with companies is beneficial for both the students who can make better
designs based on engineering approaches at a professional level, and
for the companies who can hire better-prepared students with real-world
experience.

\section{Acknowledgments}

On behalf of the Oxford Brookes Racing team, the authors wish to thank
all sponsors who supported and believed in the ambitious mission of
designing a 4WD electric vehicle. In addition, we would like to thank
all the professionals from industry, the OBR team members and the
alumni involved in this work, their valuable feedback was fundamental
in achieving the aims of this paper. Finally, we would like to thank
Oxford Brookes University for its financial support, and to the university
staff for their support, advice and belief in the team.

\bibliographystyle{unsrtnat}
\bibliography{references}

\end{document}